\documentclass[conference]{IEEEtran}
\IEEEoverridecommandlockouts
\usepackage{cite}
\usepackage{amsmath,amssymb,amsfonts}
\usepackage{algorithmic}
\usepackage[pdftex]{graphicx}
\usepackage{textcomp}
\usepackage[pdftex]{xcolor}
\usepackage{caption}
\usepackage{url}
\usepackage{subcaption}
\usepackage{booktabs,ragged2e}
\usepackage{fancyhdr}
\def\BibTeX{{\rm B\kern-.05em{\sc i\kern-.025em b}\kern-.08em
    T\kern-.1667em\lower.7ex\hbox{E}\kern-.125emX}}
\usepackage{fancyhdr}
\begin{document}
\bstctlcite{IEEEexample:BSTcontrol}

\newcommand{\TODO}[1]{\color{red}\textbf{[TODO: #1]}\color{black}}
\newcommand{\NOTICE}[1]{\color{blue}\textbf{[NOTICE: #1]}\color{black}}
\newcommand{\REFER}{\color{magenta}\textbf{[NEED REFERENCE]}\color{black}}
\newcommand{\etal}{{\it et al.}}

\captionsetup[subfigure]{labelformat=simple}
\renewcommand{\thesubfigure}{(\alph{subfigure})}

\title{Enhancing Model Learning and Interpretation using Multiple Molecular Graph Representations for Compound Property and Activity Prediction}
\author{\IEEEauthorblockN{%
% 1\textsuperscript{st} 
Apakorn Kengkanna}
\IEEEauthorblockA{\textit{Department of Computer Science} \\
\textit{School of Computing}\\
\textit{Tokyo Institute of Technology}\\
Kanagawa, Japan\\
kengkanna@li.c.titech.ac.jp}
\and
\IEEEauthorblockN{%
% 2\textsuperscript{st} 
Masahito Ohue}
\IEEEauthorblockA{\textit{Department of Computer Science} \\
\textit{School of Computing}\\
\textit{Tokyo Institute of Technology}\\
Kanagawa, Japan\\
ohue@c.titech.ac.jp}
}

\maketitle
%copyright notice
\thispagestyle{plain}
 \fancypagestyle{plain}{
 \fancyhf{} % clear all header and footer fields
%  \fancyfoot[L]{  \vspace{-8mm}\small
%978-1-6654-8462-6/22/\$31.00~\copyright2022~IEEE \\2022 IEEE Conference on Computational Intelligence in Bioinformatics and Computational Biology (CIBCB) ~DOI:10.1109/CIBCB55180.2022.9863032} % change copyright notice
% here if required
 \renewcommand{\headrulewidth}{0pt}
 \renewcommand{\footrulewidth}{0pt}
 }

\begin{abstract}
Graph neural networks (GNNs) demonstrate great performance in compound property and activity prediction due to their capability to efficiently learn complex molecular graph structures. However, two main limitations persist including compound representation and model interpretability. While atom-level molecular graph representations are commonly used because of their ability to capture natural topology, they may not fully express important substructures or functional groups which significantly influence molecular properties. Consequently, recent research proposes alternative representations employing reduction techniques to integrate higher-level information and leverages both representations for model learning. However, there is still a lack of study about different molecular graph representations on model learning and interpretation. Interpretability is also crucial for drug discovery as it can offer chemical insights and inspiration for optimization. Numerous studies attempt to include model interpretation to explain the rationale behind predictions, but most of them focus solely on individual prediction with little analysis of the interpretation on different molecular graph representations. This research introduces multiple molecular graph representations that incorporate higher-level information and investigates their effects on model learning and interpretation from diverse perspectives. Several experiments are conducted across a broad range of datasets and an attention mechanism is applied to identify significant features. The results indicate that combining atom graph representation with reduced molecular graph representation can yield promising model performance. Furthermore, the interpretation results can provide significant features and potential substructures consistently aligning with background knowledge. These multiple molecular graph representations and interpretation analysis can bolster model comprehension and facilitate relevant applications in drug discovery. 

\end{abstract}


%%%%%%%%%%%%%%%%%%%%%%%%%%%%%%%%%%%%%%
\section{Introduction}
%%%%%%%%%%%%%%%%%%%%%%%%%%%%%%%%%%%%%%
\begin{figure*}[htbp]
    \centering
    \includegraphics[width=\linewidth]{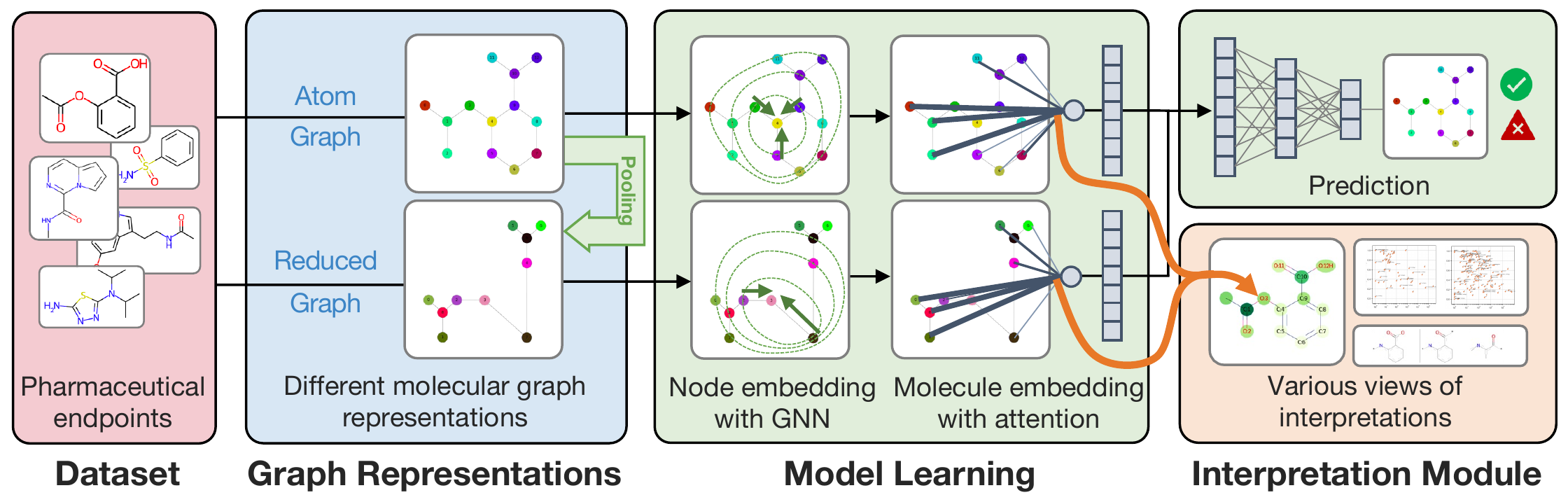}
    \caption{Research workflow}
    \label{fig:workflow}
\end{figure*}

Drug discovery processes have been driven by various advanced artificial intelligence techniques, particularly chemical properties and activity prediction. Such methods can accelerate the research processes by providing high-throughput results, decreasing cost, saving time before conducting biochemical experiments~\cite{Dara2021}. Graph neural network (GNN) is a potential technique that is gaining popularity in this field because of its exceptional performance and ability to learn natural and expressive features from molecular graph representation using node and edge relationships~\cite{Gaudelet2021}. However, there are two main important issues for GNNs, including the way to represent the compounds and the interpretability of the model.

In general, most research represents the compounds using atom-level molecular graph representation, in which nodes and edges represent atoms and bonds, respectively. This representation provides a natural structure of the molecule and has been used commonly in many applications~\cite{Tang2020,Yang_Z2022}. Although atom-level molecular graph representation has the advantage of capturing all elements and topology of compound, it has limitations due to a lack of information about substructures, containing functional groups or pharmacophore features, which play important roles in identifying property and interaction of compound~\cite{Jin2021}. To capture large substructures, the learning depth of GNNs should be increased, which might cause an over-smoothing problem when the nodes end up with similar embeddings~\cite{Ji2022}. Moreover, due to the detailed representation, the interpretation results are sometimes scattered and inconsistent within the same substructure~\cite{Harren2022}. Therefore, recent research suggests alternative ways to represent compounds using the reduction techniques. This technique constructs the abstract molecular graph by collapsing a group of atoms into a single node and encoding it with higher-level information using arbitrary or knowledge rules, such as frequently-occurring substructures, predefined functional groups, or pharmacophoric features, while preserving the topological structure~\cite{Birchall2011,Kong2022}. There are several types of reduced molecular graph representations that vary in their level of specificity and degree of aggregation. Many studies take advantage of both atom-level molecular graph representation and reduced molecular graph representation by embedding the compounds with both representations and show great performance in various applications~\cite{Jin2021,Nakano2021,Ye2022}. However, different reduced molecular graphs may not be applicable for all tasks, and there is still less research and analysis on how to effectively integrate multiple molecular graph representations to support model learning and interpretation.

Another key issue with GNNs is their interpretability. GNNs can be considered black-box models since they are complex and less interpretable by design. Low interpretability restricts ability to understand the reasoning behind the prediction, hinders model improvement, and can lead to low acceptance from users in some domains. Therefore, interpretation techniques have been introduced to address these issues by rationalizing the prediction results and producing more transparent models for humans.~\cite{Arrieta2020,Oviedo2022}. In drug discovery, interpretation promotes several advantages, for instance, capturing important chemical features, extracting underlying scientific insights, and offering guidance for the next optimization process. Moreover, interpretation allows model developers to debug and avoid bias. On top of that, the explainable results can enable safety measures, increase confidence, and build trust~\cite{Arrieta2020,Lou2022}. Several works have studied the explanation of GNNs~\cite{JimnezLuna2020,Weber2021}, especially the attention-based mechanism~\cite{Ye2022,Lou2022,Wang2022}. However, most attention-based interpretations provide the explanation only on a certain view of a single prediction, which may not fully imply the overall learning of the model. In addition, there is a lack of research on the interpretation analysis of different molecular graph representations on real pharmaceutical endpoint datasets. 

Based on these limitations, this research aims to investigate multiple molecular graph representations with GNNs for molecular property/activity prediction, and to provide chemically meaningful explanations in various views to improve interpretability with appropriate evaluation. The research workflow is outlined in Fig.~\ref{fig:workflow}. The major contributions are:
\begin{itemize}
    \item Introducing and performing analysis on different molecular graph representations using reduction techniques on different levels of information.
    \item Incorporating multiple molecular graph representations into model learning and performing several extensive experiments on the relevant datasets.
    \item Integrating an interpretation module using an attention-based mechanism to extract significant features and presenting the explainable results in many viewpoints together with the evaluation with background knowledge to enhance model understanding and interpretability.
\end{itemize}

%%%%%%%%%%%%%%%%%%%%%%%%%%%%%%%%%%%%%%
\section{Materials and Methods}
%%%%%%%%%%%%%%%%%%%%%%%%%%%%%%%%%%%%%%
\subsection{Molecular Graph Representations}
\begin{table*}[tb]
\centering
\caption{Features of molecular graph representations}
   \scalebox{1.0}{
        \begin{tabular}{lccp{5cm}p{3.5cm}}
        \toprule
        \textbf{Graph}         & \textbf{\#Node features}     & \textbf{\#Edge Features}              & \textbf{Node Features}                & \textbf{Edge Features}               \\ \midrule
        Atom (A)                   & 79                & 10                & Atom properties                & Bond Properties                \\
        Pharmacophore (P) & 6                 & 3                & Predefined node types from\newline Extended Reduced Graphs (ErG)                 & Types of connected nodes                \\
        JunctionTree (J) & 83 & 6                 & Edges, rings, and intersection atoms\newline with number of each atom                 & Types of connected nodes                \\
        FunctionalGroup (F) & 115                & 20                & Predefined types of edges, rings,\newline and functional groups     & Types of connected nodes\newline and number of intersections                \\
        \bottomrule
        \end{tabular}
    }
    \label{tab:molgraphrep}
\end{table*}

\begin{figure*}[tb]
    \centering
    \includegraphics[width=.95\linewidth]{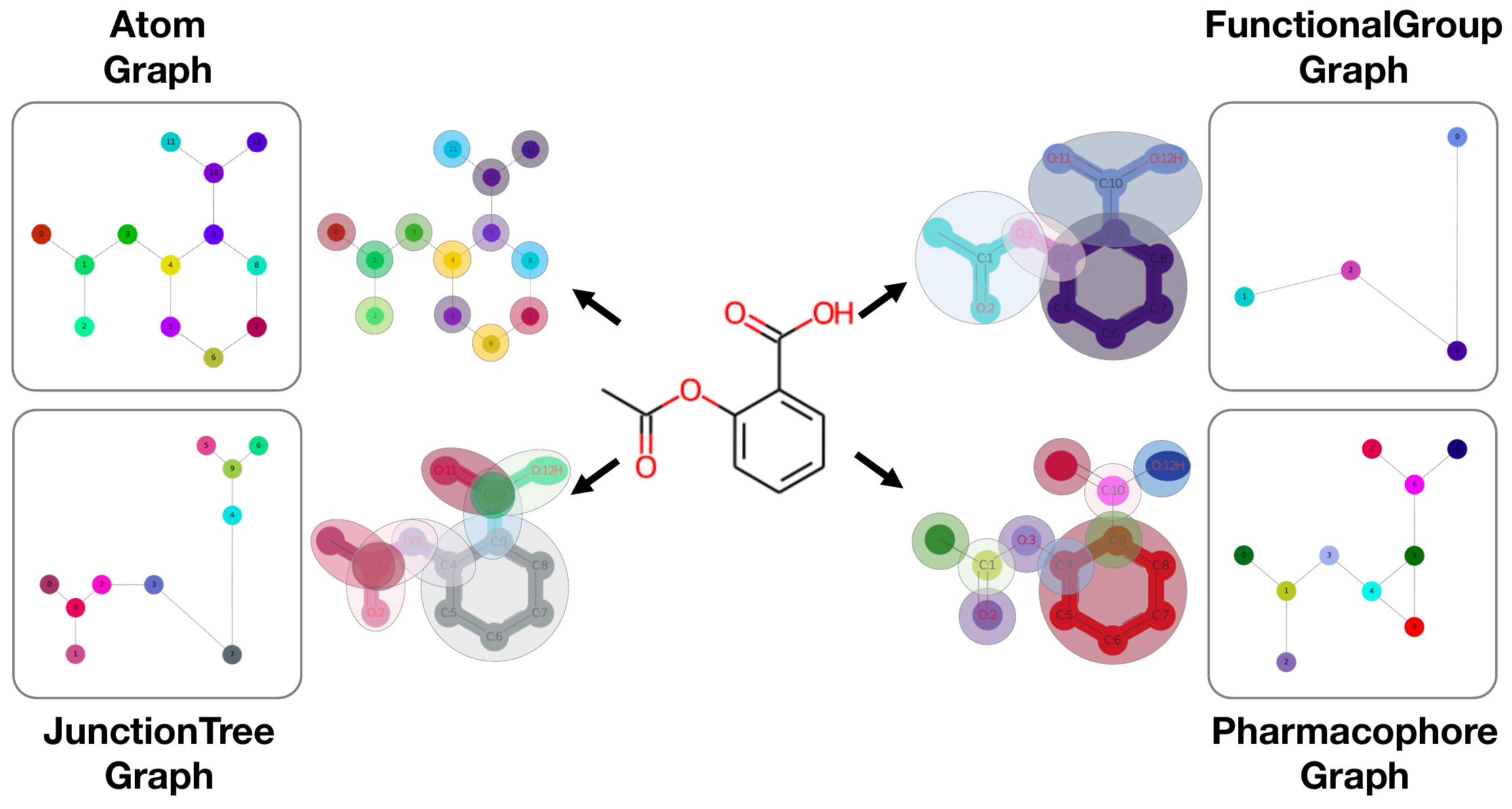}
    \caption{Different molecular graph representations of aspirin}
    \label{fig:molgraph}
\end{figure*}

A molecule can be depicted as a graph structure to describe its chemical topology in term of the relationships between nodes and edges. The features can be constructed using nodes, edges, and an adjacency matrix. With those features, a graph or a subgraph can hold interpretable meaning related to chemical knowledge~\cite{David2020}. Different graph topologies are available to visualize the molecule in a specific level of information. In this study, four distinct molecular graph representations are investigated as shown in Fig.~\ref{fig:molgraph}. The details and characteristics of each representation are described below and the summary of node and edges features is organized in Table~\ref{tab:molgraphrep}.
\begin{enumerate}
\item {\bf Atom graph (A)} is the most common and simplest representation in which an atom is represented as a node and a bond is represented as an edge. This representation presents the compound in a natural form and maintains all topological information including substituent positions. The node and edge properties are typically derived from the properties of the atom and bond. There are many applications using atom graph, including molecular property prediction and drug-target interaction~\cite{Jin2021,Xiong2020}. The drawbacks are the lack of substructure information, over-smoothing problems when increasing model depth, and sparse interpretation.
\item {\bf Pharmacophore graph (P)} represents node features associated with binding activity and pharmacophoric information using the extended reduced graphs (ErG) algorithm~\cite{Stiefl2006}. The node is encoded by the combination of six features including H-bond donor, H-bond acceptor, positive, negative, hydrophobic, and aromatic. This graph shows great performance in scaffold hopping and protein-ligand interaction tasks~\cite{Nakano2021,Kong2022}, but the representation is limited to only six predetermined node types for this specific algorithm. 
\item {\bf JunctionTree graph (J)} is widely used in molecule generation tasks~\cite{Jin_J2018}. The original atom graph is transformed into a tree-based structure by converting bonds, rings, and junction atoms into nodes. This no-loop structure gives an advantage to message-passing learning by reducing the dead-loop problem and repeated information issues~\cite{Jin2021,Ye2022}. By the way, this representation also does not include functional groups and ring types, and there are difficulties in forming complex rings.
\item {\bf FunctionalGroup graph (F)} is the representation that integrates functional group information by changing a subgraph into a single node using predefined functional groups, ring types, and atom pairs~\cite{Ji2022,Lukashina2020,Kwon2020}. It allows higher-level understanding of node features with chemical background. However, this representation still has limitations in representing complicated rings and functional groups that cannot be fully predefined.
\end{enumerate}

\subsection{Graph-based Model Interpretation}
An attention mechanism is employed in this study to extract the interpretation from the model. The attention mechanism implicitly assigns the attention weights to the nodes so that more important nodes will receive higher weights during neighborhood aggregation~\cite{Velickovic2018}. Attention mechanism is introduced inside the model and learned during the model training to improve performance. Therefore, explaining attention weights would be useful for understanding model learning.

To visualize the interpretation, most of the research focuses on local-level explanation of a single prediction which may not convey the overall reasoning of model. This research proposes three different views of interpretation to provide a comprehensive understanding and new insights. Different evaluation methods are used to verify the interpretation of different views.
\begin{enumerate}
\item {\bf Single prediction view} illustrates the interpretation of individual molecule prediction and examines the results with background knowledge. For the ligand binding activity task, this interpretation can be compared with the binding mode observed from the interaction map of complex structures. This view shows the specific important portion of a single molecule, which can inspire compound optimization and simplification task.
\item {\bf Node features view} visualizes interpretation in collections of node features presented in the specific plot. The evaluation can be analyzed with general background knowledge. This view provides a general understanding of collection of molecules in dataset for trend analysis.
 \item {\bf Potential substructures view} analyzes interpretation in the format of structural patterns to provide chemically intuitive understanding. The molecules are fragmented into substructures and the substructures with outstandingly high attention weights based on detection rules are collected and defined as potential substructures. These potential substructures are then assessed with the key structural patterns reported in the literature for evaluation. This view benefits for suggesting interesting structural modification or subsequent optimization.
\end{enumerate}

\section{Experiments}
%%%%%%%%%%%%%%%%%%%%%%%%%%%%%%%%%%%%%%
\subsection{Datasets}
This study utilizes two sets of datasets to validate model performance and interpretation. One dataset comes from MoleculeNet~\cite{Wu2018}, which are the general benchmarks widely used in the field of compound property prediction. Another dataset contains seven pharmaceutical endpoints obtained from various sources. These datasets are important for validating intepretation because they have been extensively reviewed and reported about task-related substructures that are relevant to specific properties and activities, so-called key structural patterns. This research aims to validate the interpretation results based on the assumption that the interpretation results should be able to identify trends and potential substructures corresponding with those key structural patterns. Table~\ref{tab:dataset} summarizes all datasets in this study. 
\begin{table}[tb]
    \centering
    \caption{Datasets statistics}
    \scalebox{1.0}{
        \begin{tabular}{lccc}
        \toprule
        \textbf{Dataset}               & \textbf{Task}  & \textbf{\begin{tabular}[c]{@{}c@{}}\#{}Compounds\end{tabular}} & \textbf{\begin{tabular}[c]{@{}c@{}}Positive/\\Negative\end{tabular}} \\ \midrule
\multicolumn{4}{l}{\textbf{Benchmarks from MoleculeNet}}                          \\
BACE~\cite{Wu2018}      & Classification & 1,513               & 1.19                         \\
BBBP~\cite{Wu2018}      & Classification & 2,050               & 0.31                         \\
FreeSolv~\cite{Wu2018} & Regression & 642               & -                         \\
ESOL~\cite{Wu2018}              & Regression & 1,128               & -                         \\
Lipo~\cite{Wu2018}            & Regression & 4,200                 & -                         \\
\multicolumn{4}{l}{\textbf{Pharmaceutical Endpoints}}                          \\
AmesMutag~\cite{Hansen2009} & Classification & 6,512               & 1.16                         \\
hERG20~\cite{Cai2019}              & Classification & 6,548               & 1.99                         \\
CYP2C8~\cite{Zhang2021}            & Classification & 553                 & 1.41                         \\
Hepatotoxicity~\cite{He2019}      & Classification & 1,489               & 1.12                         \\
HumanPPB~\cite{Lou2022}            & Regression     & 3,921               & -                            \\ 
AqSolDB~\cite{Sorkun2019}          & Regression     & 9,982               & -                            \\ 
HIV1~\cite{Li_Y2018}          & Regression     & 2,602               & -                            \\ 
        \bottomrule
        \end{tabular}
    }
    \label{tab:dataset}
\end{table}

\subsection{Model Architecture}
\begin{figure*}[t]
    \centering
    \includegraphics[width=\linewidth]{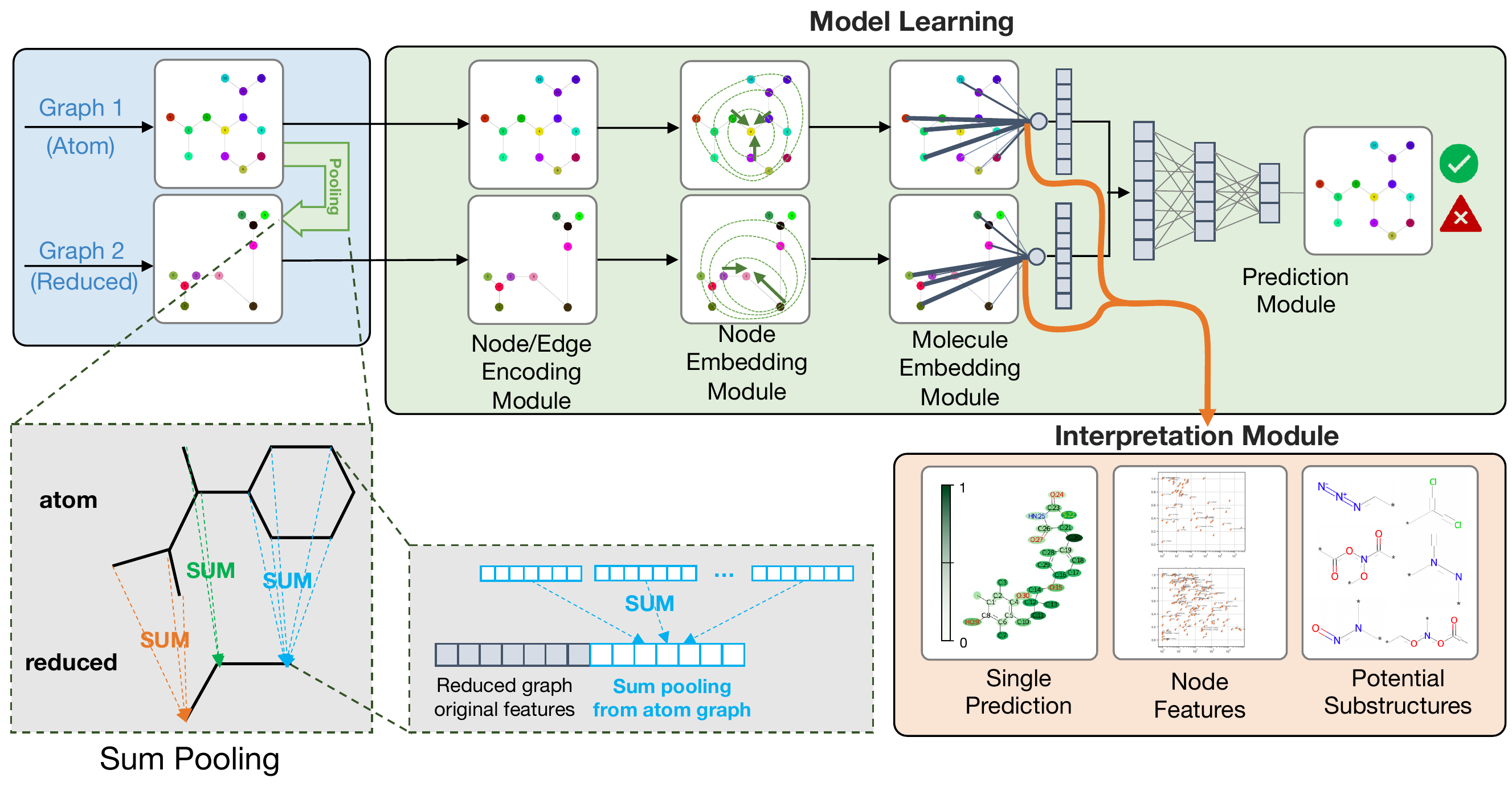}\vspace{-3mm}
    \caption{Illustration of model architecture and interpretation}
    \label{fig:schema}
\end{figure*}

The model is designed and inspired by AttentiveFP~\cite{Xiong2020} and its variants~\cite{Yang2022,Tang2020}. There are 4 components for GNN learning as shown in Fig.~\ref{fig:schema}. Firstly, the node and edge encoding module encodes the initial node and edge features from each molecular graph representation into the fixed-size vectors. Secondly, the node embedding module is implemented to learn node and edge features by using modified graph isomorphism network (GIN) that takes edge features in neighboring aggregation~\cite{Hu_W2019}. This step integrates the use of gate recurrent units (GRUs) before updating each node because GRUs show good ability in controlling information to be aggregated or reserved from neighboring nodes~\cite{Yang2022}. The third component is a molecule embedding module with an attention mechanism. In this step, the concept of a virtual super-node connecting all nodes in the graph is introduced to readout features from all node embeddings using a graph attention network (GAT). This component also integrates the GRUs to retain and filter pooled features resulting in molecule embedding. Importantly, the GAT readout in this module provides attention weights for each node in the graph which will be used as the interpretation. The nodes that receive high attention weights are assumed to be important for that particular prediction. The last component is the prediction module which combines molecule embeddings from multiple graphs and performs classification or regression using fully connected layers accordingly.

\subsection{Feature and Experiment Setup}
This study conducts experiments on various datasets with several graph representations. There are 4 experimental schemes, including Atom graph (A) and the combinations of the atom graph with another reduced molecular graph, which are Pharmacophore graph (A+P), FunctionalGroup graph (A+F), and JunctionTree graph (A+J). For the scheme of combining two molecular graph representations, there is a special process to initialize reduced molecular graph node features. Apart from the original node features, the node features of the reduced molecular graph are enriched by expanding with the pooled node features from the corresponding nodes in the atom graph using sum-pooling as shown in Fig.~\ref{fig:schema}.

\subsection{Training and Hyperparameter Tuning}
The datasets are divided into train, validation and test sets in the ratio of $8:1:1$. The splitting method is suggested based on the original paper if specified; otherwise, random splitting is used. The models are trained using 5-fold cross-validation, and hyperparameter tuning is performed using Optuna library~\cite{Akiba2019}. The major hyperparameters are batch size, dimensions of hidden layers, and number of node and molecule embedding layers for atom graph and reduced graph. The molecule embedding size is set to 256 dimensions. The learning rate, weight decay, dropout rate and batch normalization are set appropriately for each dataset. AUROC and RMSE are used as performance measures for classification and regression tasks, respectively. All models are trained for 300 epochs, but training is stopped if the performance of the validation set is not improved for 30 epochs. 

\subsection{Interpretation Extraction}
The interpretation of the model is extracted from the attention mechanism during the molecule embedding module. For each graph, the attention weights at the edges connected to a virtual super-node are extracted and normalized using the min-max algorithm. For a reduced molecular graph, the interpretation can be visualized using the node features directly or using the mapping function method to the original atom graph. For the later method, the attention weights at the nodes are mapped back to the corresponding nodes in the original atom graph. Then, the min-max normalization is also applied. In the case of multiple graphs, after the attention weights are mapped back to the original atom graph, the attention weights are merged by selecting the max attention weights among all graphs to give high priority to the focused part of the molecule. The extracted attention weights are visualized on the different views to provide an intensive understanding of the model.

For the potential substructures view, the potential substructures are extracted using the following procedures. Firstly, the compounds are fragmented using fragmentation methods including BRICS~\cite{Degen2008}, RECAP ~\cite{Lewell1998}, and GRINDER~\cite{Lou2022} to get all possible fragments with 3--20 atoms. The fragments having a median of attention weights greater than $75^{th}$ percentile of molecule attention weights are labeled as important fragments. Then, calculate the important fragment percentage between compounds having those fragments and compounds having those fragments which are labeled as important. Next, the fragment importance score followed from~\cite{Jian2022} is used to quantify the significance of the fragments. Basically, the fragment importance score is calculated by the summation of the difference between the average attention weights of fragment and the average attention weights of the compound divided by the number of fragments. Finally, the fragment is considered as potential substructure if it passes all of these conditions: 1) The important fragment percentage is greater 50\%. 2) The number of compounds having that fragment which is labeled as important is large enough for each dataset. 3) The fragment importance score is greater than zero. After removing redundancies, the final potential substructures are analyzed and evaluated with the key structural patterns reported in the literature.

\section{Results and Discussion}
%%%%%%%%%%%%%%%%%%%%%%%%%%%%%%%%%%%%%% 
\subsection{Model Performance}
The performance of the models is presented in Table~\ref{tab:performance_class} and Table~\ref{tab:performance_reg}. The results show that combining the atom graph with a reduced molecular graph often leads to moderately better performance compared to using the atom graph alone. Notably, FunctionalGroup and Pharmacophore Graph combinations show promise based on their average ranking for all datasets. Even though different graph representations produce acceptable performance, they are inconsistent across each dataset. According to the characteristics of each dataset, certain combinations can positively facilitate model learning by providing meaningful features. On the other hand, some combinations can negatively affect the performance by introducing irrelevant features, causing bias, and increasing complexity. Therefore, the selection of molecular graph combinations and additional feature engineering should be appropriately considered depending on the nature of the datasets.

\begin{table*}[tb]
\centering
\caption{Model performance of test datasets on classification task (AUROC)}
    \scalebox{1.0}{
    \begin{tabular}{ccccccc}
    \toprule
\textbf{Model} & \textbf{BACE}         & \textbf{BBBP}         & \textbf{AmesMutag}   & \textbf{hERG20}      & \textbf{CYP2C8}       & \textbf{Hepatotoxicity} \\ \midrule
A              & 0.7090 (0.0245)       & 0.8828 (0.0187)       & 0.8604 (0.0056) & 0.9227 (0.0057)       & {\underline {0.8553 (0.0088)}} & 0.7184 (0.0087)         \\
A+F            & 0.7320 (0.0279) & 0.8859 (0.0032) & {\underline {0.8680 (0.0040)}}       & {\underline {0.9275 (0.0043)}} & 0.8505 (0.0146)       & {\underline {0.7724 (0.0131)}}   \\
A+P            & 0.7412 (0.0175)       & {\underline {0.8922 (0.0056)}}       & 0.8649 (0.0033)       & 0.9233 (0.0067)       & 0.8506 (0.0202)       & 0.7470 (0.0118)         \\
A+J            & {\underline {0.7483 (0.0323)}}       & 0.8747 (0.0077)       & 0.8611 (0.0019)       & 0.9231 (0.0048)       & 0.8443 (0.0101)       & 0.7322 (0.0187)         \\ 

    \bottomrule
    \multicolumn{7}{l}{The underlined numbers are the best performance of each dataset. The numbers in parentheses are the standard deviations.}
    \end{tabular}
    }
    \label{tab:performance_class}
\end{table*}

\begin{table*}[tb]
\centering
\caption{Model performance of test dataset on regression task (RMSE)}
    \scalebox{1.0}{
    \begin{tabular}{ccccccc}
    \toprule
\textbf{Model} & \textbf{FreeSolv}      & \textbf{ESOL}          & \textbf{Lipo}          & \textbf{HumanPPB}      & \textbf{AqSolDB}       & \textbf{HIV1}          \\ \midrule
A              & 1.5053 (0.0731)       & 0.7276 (0.0488)       & 0.5839 (0.0139)       & 0.1347 (0.0083)       & 1.0115 (0.0261)       & 1.2817 (0.0425)       \\
A+F            & {\underline {1.2677 (0.0826)}}       & 0.7460 (0.0570) & 0.5926 (0.0128)       & 0.1389 (0.0130)       & {\underline {0.9816 (0.0185)}} & {\underline {0.9914 (0.0336)}} \\
A+P            & 1.3744 (0.0792) & {\underline {0.6761 (0.0188)}}       & 0.5915 (0.0118)       & {\underline {0.1312 (0.0140)}} & 0.9882 (0.0189)       & 1.3492 (0.1630)       \\
A+J            & 1.3772 (0.0097)       & 0.7362 (0.0307)       & {\underline {0.5765 (0.0189)}} & 0.1627 (0.0094)       & 1.0016 (0.0197)       & 1.1041 (0.0499)       \\ 

    \bottomrule
    \multicolumn{7}{l}{The underlined numbers are the best performance of each dataset. The numbers in parenthesis are the standard deviations.}
    \end{tabular}
    }
    \label{tab:performance_reg}
\end{table*}

\subsection{Interpretation Results}
To validate the interpretation results, different perspectives of interpretation from attention weights are demonstrated and compared with chemical background knowledge. These results are obtained from the entire dataset using the model that produces highest performance on validation set.

\begin{figure*}[tb]
    \centering
    \begin{subfigure}[htbp]{\linewidth}
        \centering
		\includegraphics[width=0.75\linewidth]{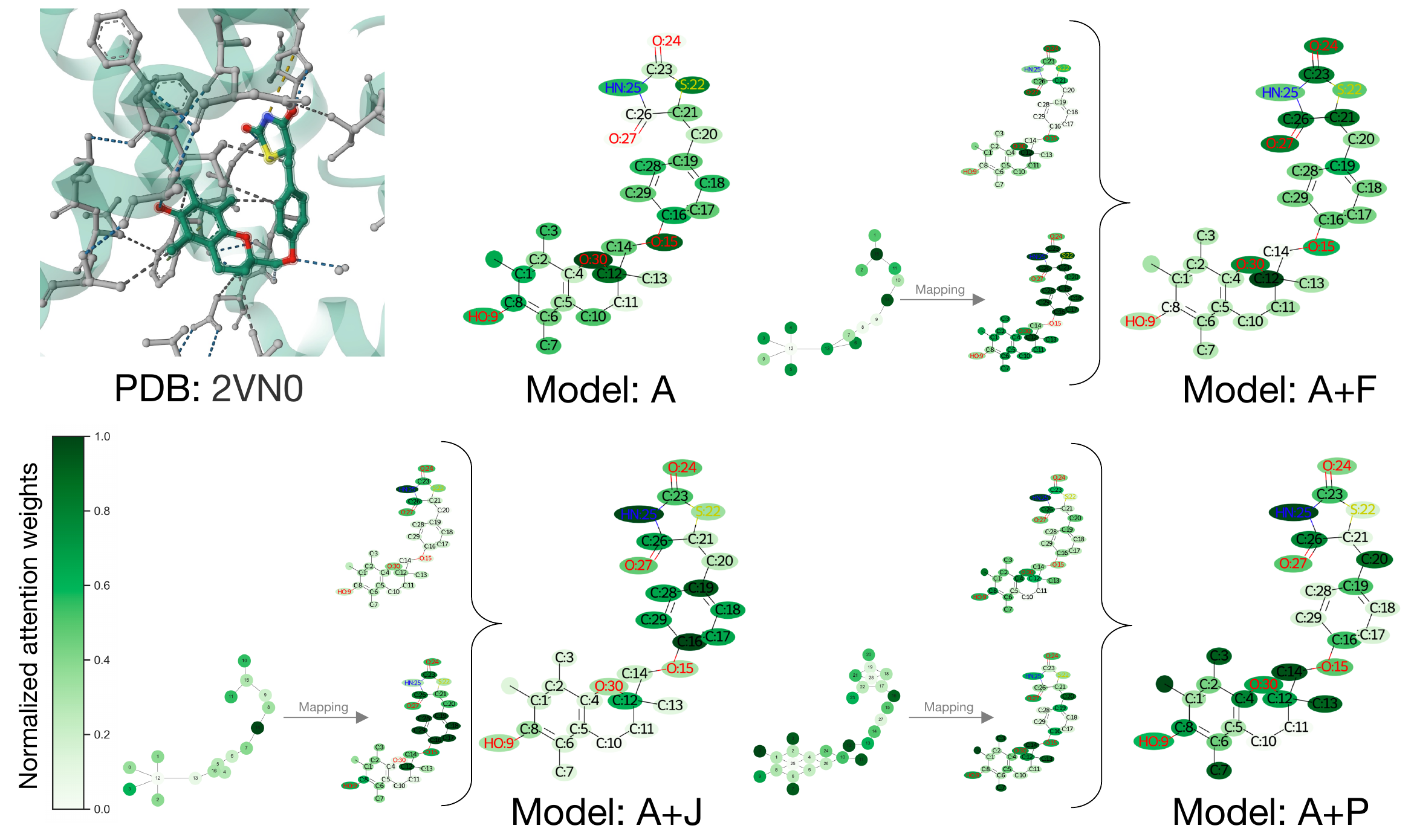}
        \subcaption[short for lof]{Troglitazone for CYP2C8 Dataset}
		\label{fig:prediction_CYP2C8}
    \end{subfigure}
    \begin{subfigure}[tb]{\linewidth}
        \centering
		\includegraphics[width=0.75\linewidth]{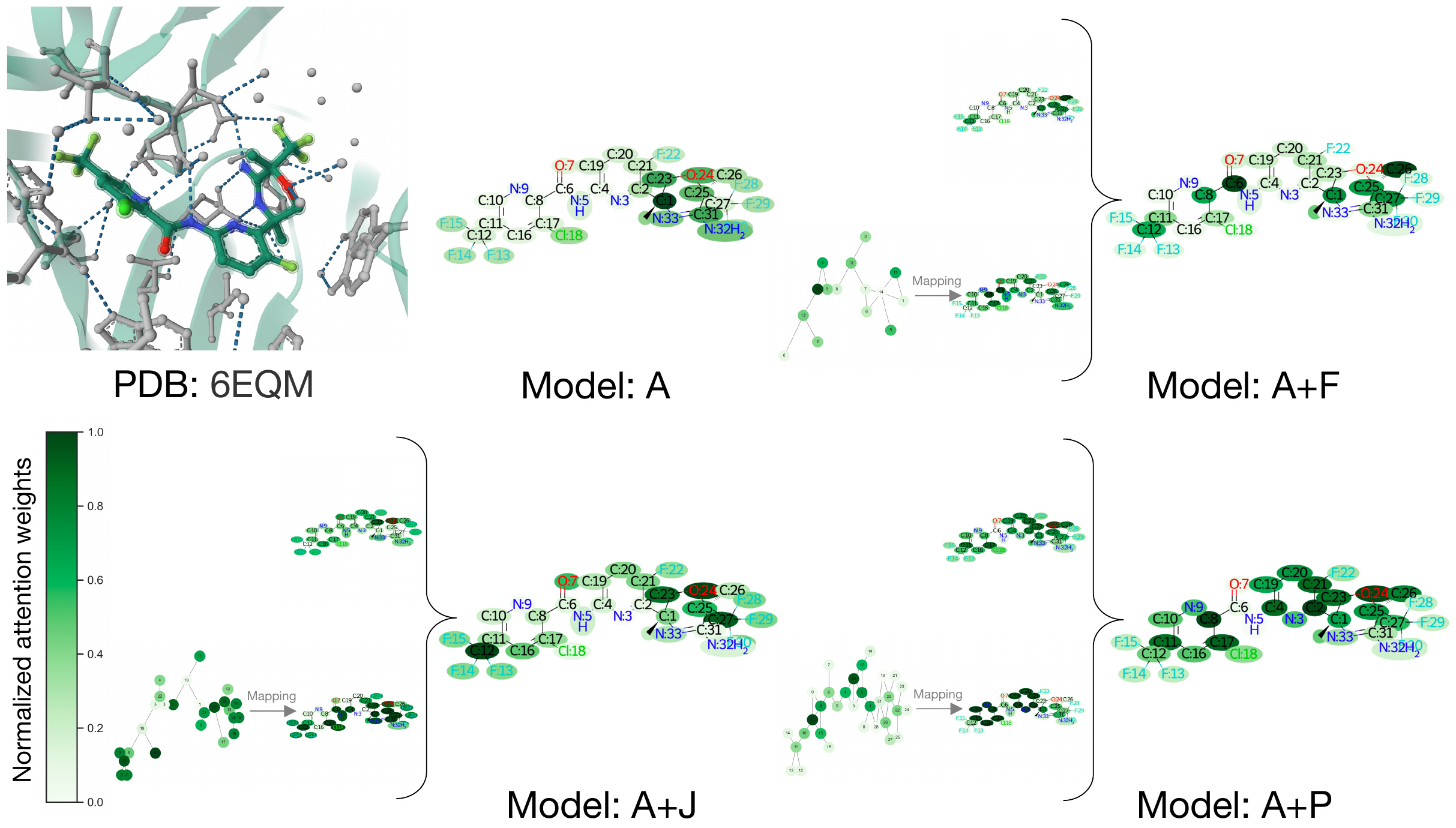}
        \subcaption[short for lof]{Umibecestat (CNP520) for BACE dataset}
		\label{fig:prediction_BACE}
    \end{subfigure}
    \hfill
    \caption{Interpretation on single prediction view from attention weights. For the multiple graphs scheme, the mapping and combination function are applied to visualize on original atom graph. (a) Troglitazone and the interaction map with CYP2C8 complex (PDB:2VN0)  (b) Umibecestat (CNP520) and the interaction map with BACE-1 complex (PDB:6EQM)}
    \label{fig:training}
\end{figure*}

Starting with the single prediction view, taking CYP2C8 dataset as the first example, the troglitazone molecule has been reported as a significant inhibitor for the CYP2C8 target~\cite{Zhang2021}. The attention weights compared with the interaction map of the complex PDB:2VN0 are shown in Fig.~\ref{fig:prediction_CYP2C8}. Model A, A+F and A+P are able to recognize the part of ether oxygen in the middle of the molecule, which can form a hydrogen bond with the residues using nearby water molecules. While, model A+F, A+J, and A+P can positively identify the thiazolidinedione fragment, which contains ketone oxygen that can form a hydrogen bond with target residues. These results align with the discussion in~\cite{Zhang2021}. Despite the different results from each model, their interpretations can well capture the main parts corresponding to the interaction region. 

Another example comes from BACE dataset. Umibecestat or CNP520 was discovered as a potent small molecule inhibitor of BACE1~\cite{Neumann2018}. The attention weights from the model are compared with the interaction map from PBD:6EQM, as shown in Fig.~\ref{fig:prediction_BACE}. Although the models cannot clearly capture the part of the oxazine nitrogen that mainly forms interactions with target residues, they still give high focus on the region of oxazine that contributes to the binding mode, and also put high attention weight to oxazine oxygen that accepts H-bonds from water molecules, as described in ~\cite{Machauer2021}. This implies that the interpretation can relatively capture the important binding regions for ligand activity.

\begin{figure*}[htbp]
    \centering
    \begin{subfigure}[htbp]{1\columnwidth}
        \centering
		\includegraphics[width=\columnwidth]{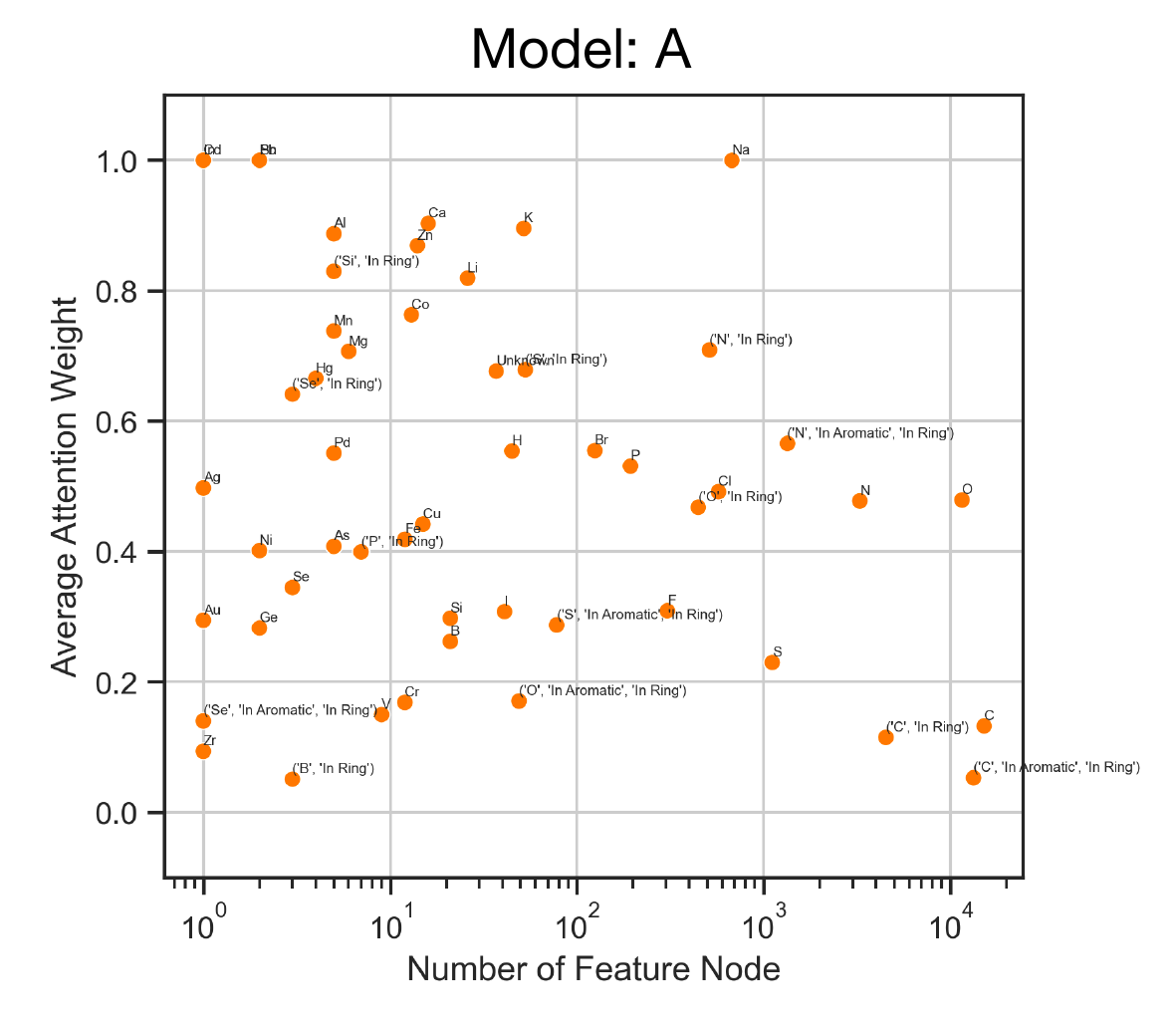}
        \subcaption{Model A}\vspace{6mm}
		\label{fig:feature_a}
    \end{subfigure}
    \hfill
    \begin{subfigure}[htbp]{1\columnwidth}
        \centering
		\includegraphics[width=\columnwidth]{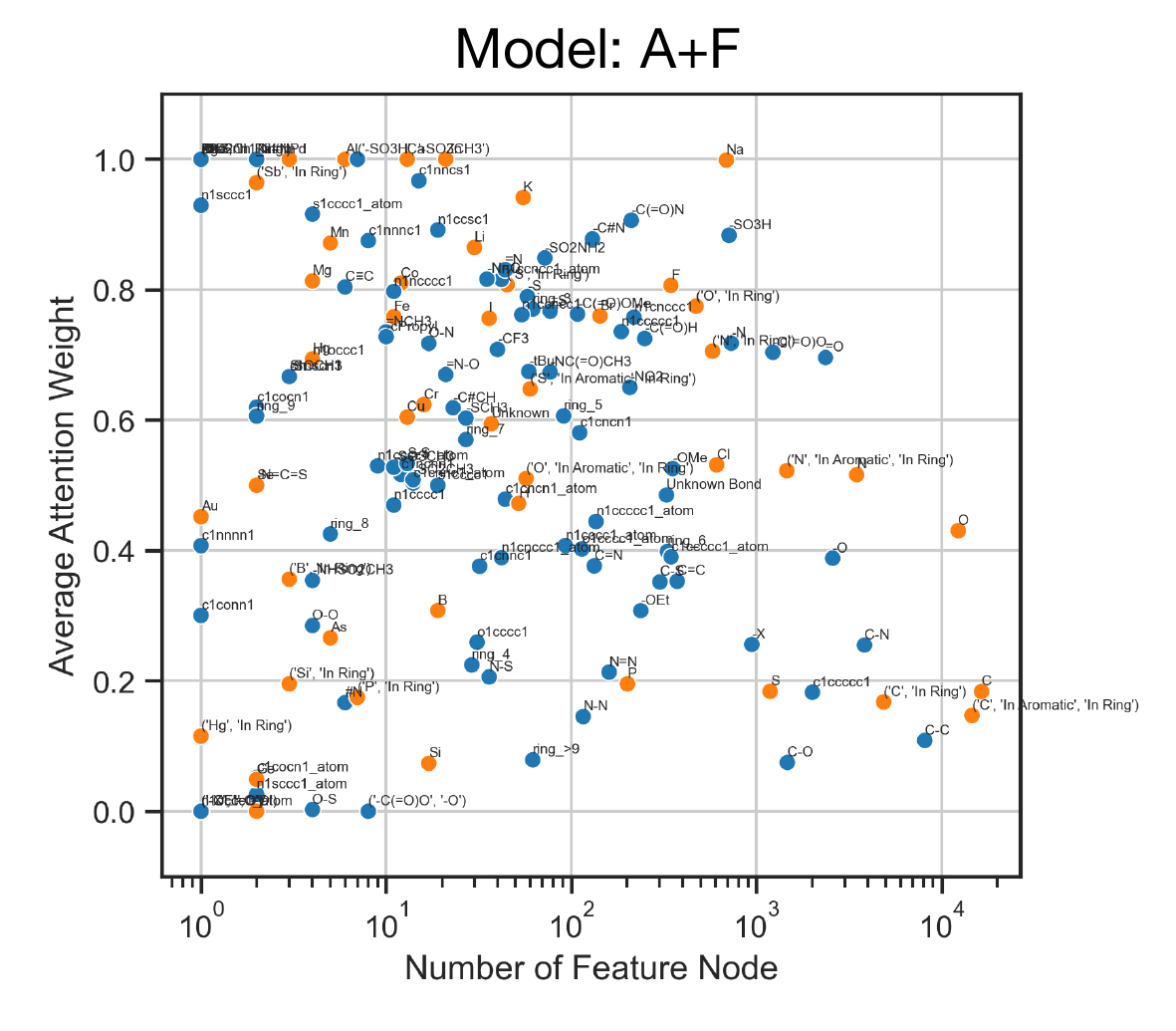}
        \subcaption{Model A+F}\vspace{6mm}
		\label{fig:feature_af}
    \end{subfigure}
    \begin{subfigure}[htbp]{1\columnwidth}
        \centering
		\includegraphics[width=\columnwidth]{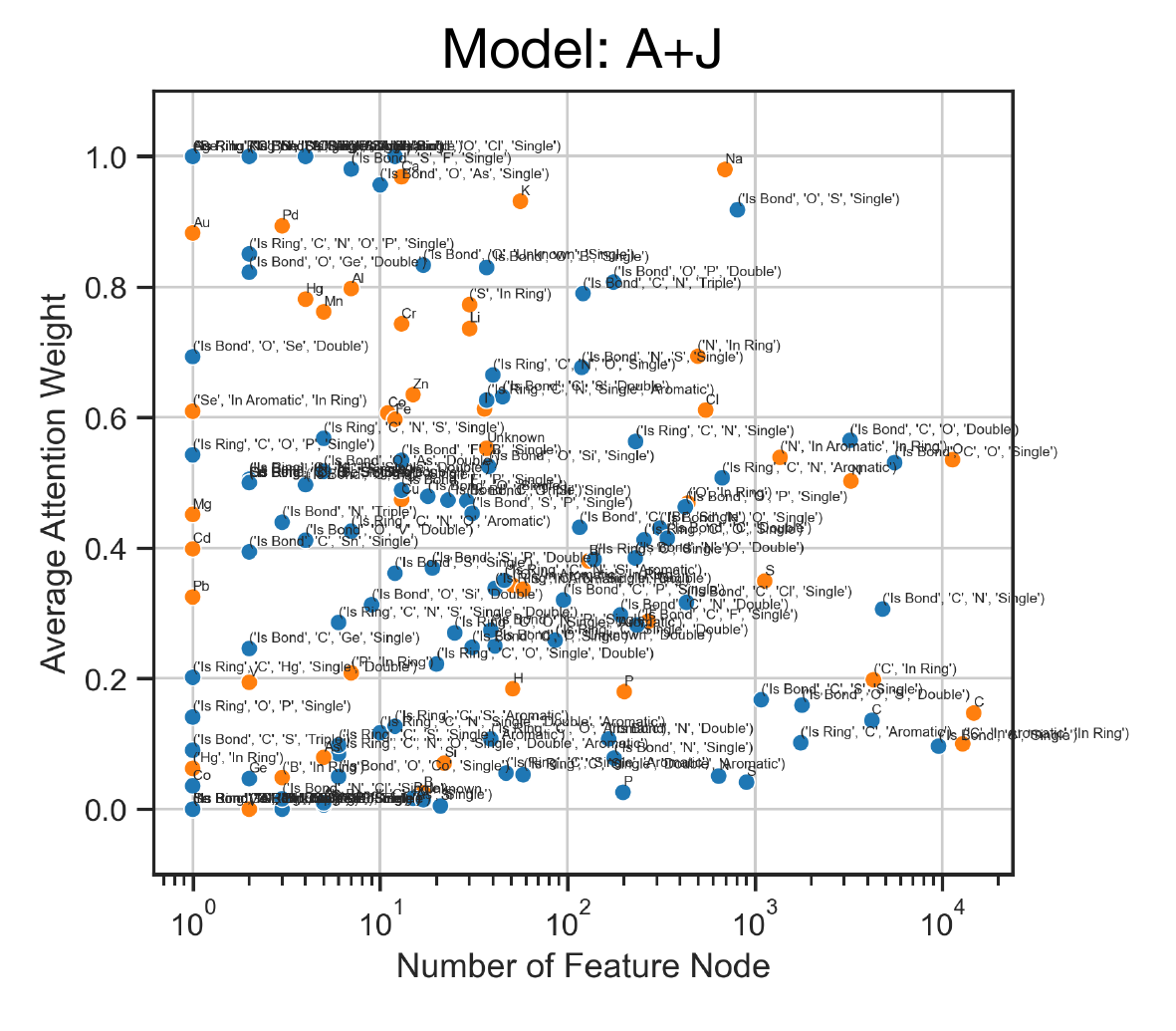}
        \subcaption{Model A+J}\vspace{3mm}
		\label{fig:feature_aj}
    \end{subfigure}
    \hfill
    \begin{subfigure}[htbp]{1\columnwidth}
        \centering
		\includegraphics[width=\columnwidth]{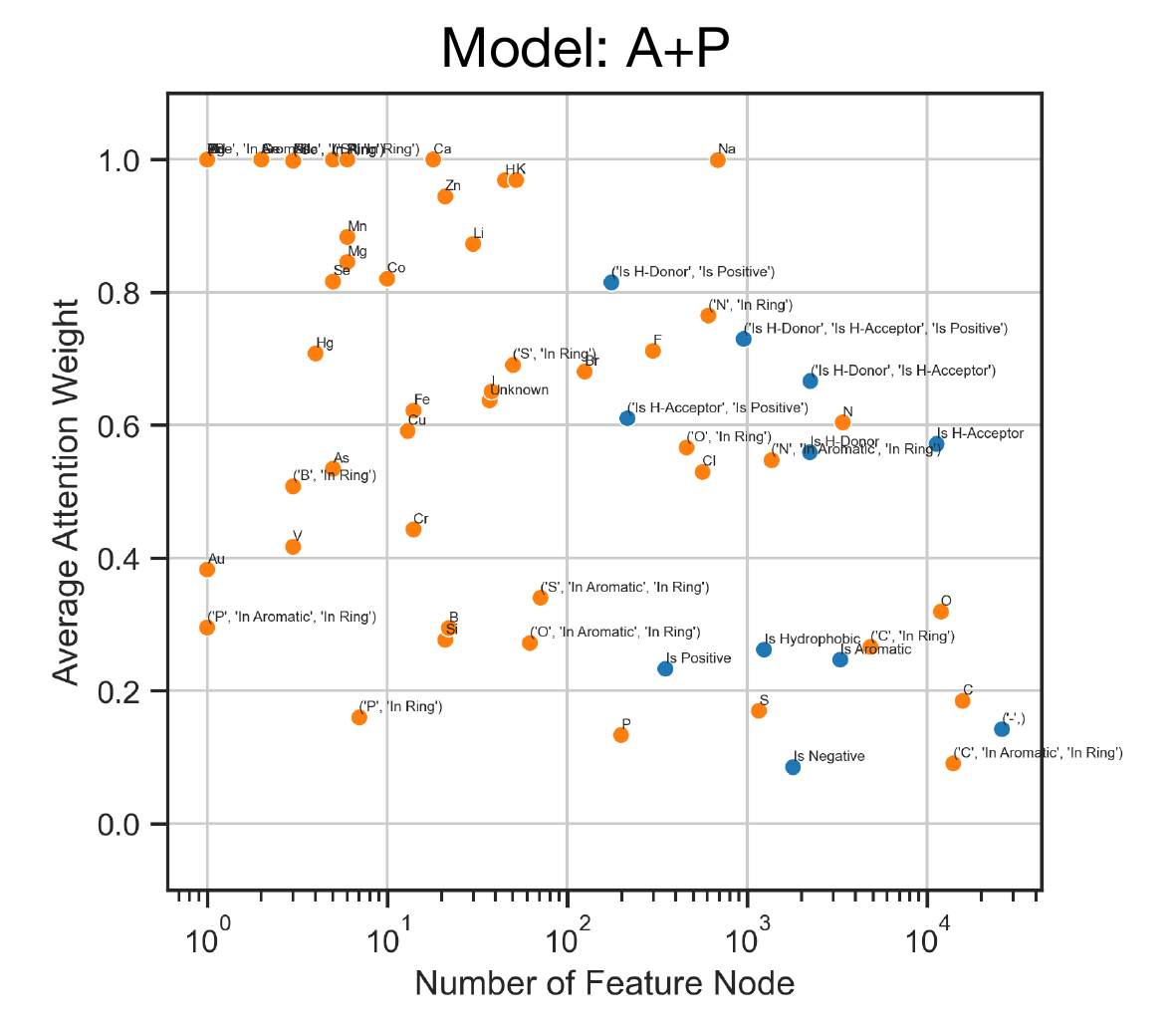}
        \subcaption{Model A+P}\vspace{3mm}
		\label{fig:feature_ap}
    \end{subfigure}
    \hfill 
    \caption{Interpretation on node features view. These graphs plot the average attention weights of each node feature with the number of feature nodes in the entire dataset. The orange dots represent the node features from the atom graph. The blue dots represent the node features from the reduced molecular graph according to the scheme.} 
    \label{fig:feature}
\end{figure*}

Next is the node features view. AqSolDB dataset is used as an example for this analysis. This dataset serves as a regression task that predicts the aqueous solubility of compounds. To visualize the significant features for a specific range of predictions, the compounds predicted as soluble and highly soluble ($\mathrm{LogS}>-2$) as classified in~\cite{Sorkun2019} are selected for analysis. The average attention weights of each node feature are recorded and plotted on the graph. As shown in Fig.~\ref{fig:feature}, the significant features can be easily observed in the area of high average attention weights and high number of feature nodes. As a result, the node features containing oxygen and nitrogen receive more importance in this case. These features are important because they are likely to form hydrogen bonds with solvents. Interestingly, the carbon atom and the aromatic ring of carbon atoms seem to attain low attention weights for all models. Evidently, the node features of combination graphs can clearly convey high-level information that is more meaningful than that of an atom graph. As such, incorporating higher-level information node features into the model can directly provide comprehensive results without requiring further processing.

\begin{figure}[tb]
    \centering
    \includegraphics[width=\columnwidth]{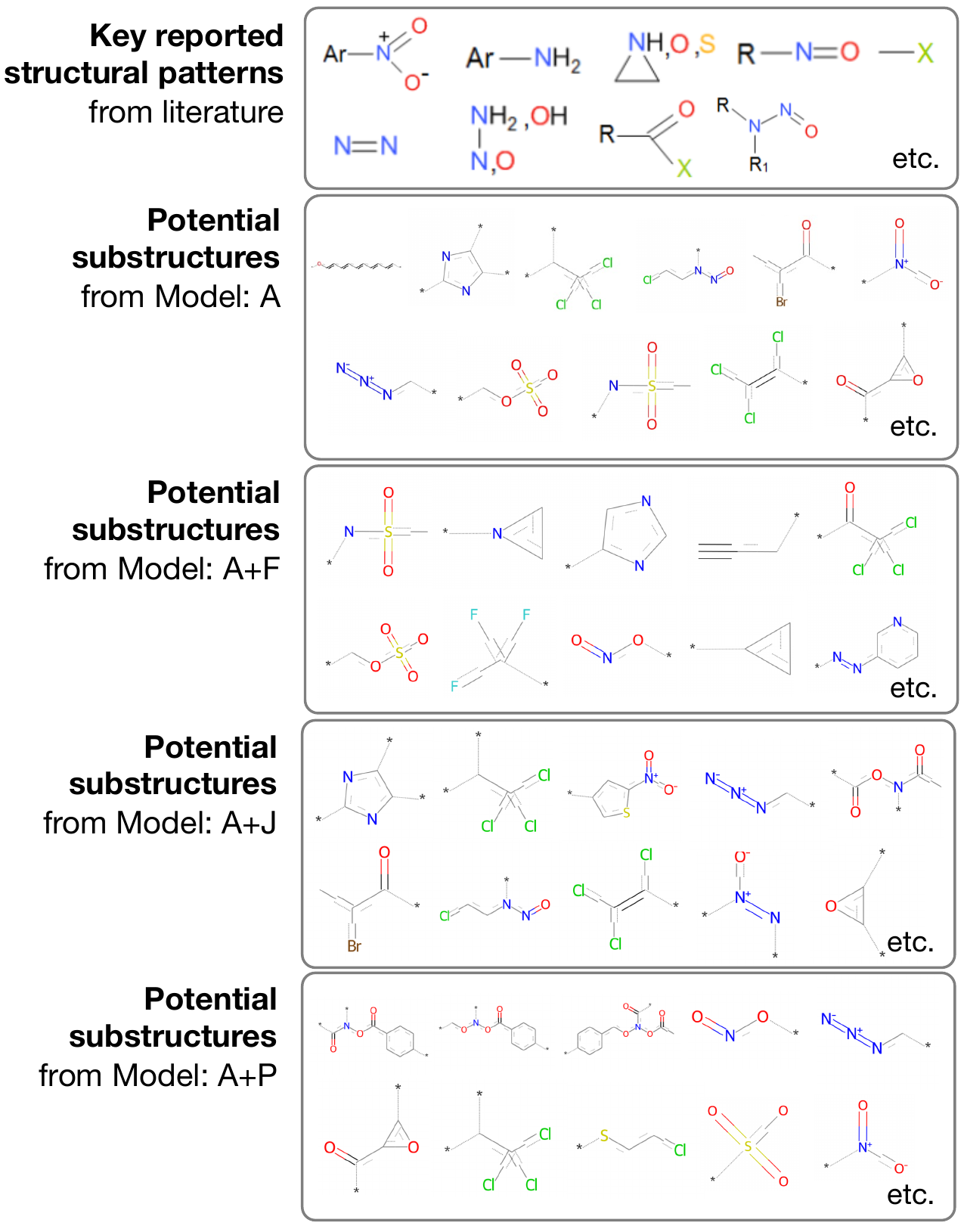}
    \caption{Interpretation on potential substructures view. Key reported structural patterns from the literature of AmesMutag dataset and potential substructures extracted from the models.}
    \label{fig:substructure}
\end{figure}

Finally, in the the potential substructures view, the molecules are fragmented into small substructures. Substructures with high attention weights are then identified as potential substructures based on detection rules. Fig.~\ref{fig:substructure} displays the potential substructures extracted from compounds predicted as positive (class 1) in AmesMutag dataset. The interpretation results can extract potential substructures that are consistent with the key structural patterns summarized in ~\cite{Sushko2012}, specifically, the fragments containing nitro, nitroso, three-membered heterocycle, and chroline. There are some interesting findings from different models; for example, sulfonate-bonded carbon atom groups are captured by models A, A+F, and A+P, which are also additional important alerts of this dataset. Model A and A+J can identify bromine, an aliphatic halide in addition to chlorine, which is also an important alert of this dataset~\cite{Sushko2012}. These results offer very useful scientific information that could be used as a guide for structural modification and other relevant tasks.

%%%%%%%%%%%%%%%%%%%%%%%%%%%%%%%%%%%%%%
\section{Conclusion}
%%%%%%%%%%%%%%%%%%%%%%%%%%%%%%%%%%%%%%
We proposed multiple molecular graph representations using graph reduction techniques to create higher-level molecular graph features. Numerous experiments had been performed on various datasets for molecular property/activity prediction. This study had shown that different molecular graph representations provided different levels of information that could support model prediction and interpretation. While combining multiple graph representations could slightly improve performance for most datasets, the performance still varied depending on the dataset. Therefore, it was crucial to carefully consider selecting graph representations and feature engineering for building prediction models. Aside from the above-mentioned molecular graph representations, there are other interesting and plausible graph representations that should have been explored in future work. These include 3D molecular graph, fragment-based molecular graph, or learned molecular graph representation, which provide a useful aspect of chemical features. Attention-based interpretations from multiple perspectives could promote a better understanding of model predictions which were also relatively consistent with chemical knowledge. These interpretations provided insightful findings that could possibly facilitate subsequent processes, such as molecular optimization and structural modification in drug discovery applications in the future.

%%%%%%%%%%%%%%%%%%%%%%%%%%%%%%%%%%%%%%
\section*{Acknowledgment}
%%%%%%%%%%%%%%%%%%%%%%%%%%%%%%%%%%%%%%
The authors thank Yutaka Akiyama and Keisuke Yanagisawa for their constructive discussion and feedback. This work was supported by JST FOREST (JPMJFR216J), JSPS KAKENHI (23H04887), and AMED BINDS (JP22ama121026).

\bibliographystyle{IEEEtran}
\bibliography{mainbib}
\end{document}